\documentclass[twocolumn]{pasj01}
\tenpoint
\Received{$\langle$reception date$\rangle$}
\Accepted{$\langle$acception date$\rangle$}
\Published{$\langle$publication date$\rangle$}
\SetRunningHead{Astronomical Society of Japan}{Usage of \texttt{pasj00.cls}}

\usepackage{times}
\usepackage[T1]{fontenc}
\usepackage{natbib}

\begin{document}
\SetRunningHead{K.A.P. Singh et al.}{Nonlinear Instability and Intermittent Nature of Magnetic Reconnection}
\Received{}%{yyyy/mm/dd}
\Accepted{}%{yyyy/mm/dd}

\title{Nonlinear Instability and Intermittent Nature of Magnetic Reconnection in Solar Chromosphere}
\author{%
K.A.P. Singh \altaffilmark{1},
Andrew Hillier \altaffilmark{2},
H. Isobe \altaffilmark{3,4},
K. Shibata \altaffilmark{4}
}
\altaffiltext{1}{UM-DAE Centre for Excellence in Basic Sciences, Mumbai 400098, India}
\email{alkendra.singh@cbs.ac.in}
\altaffiltext{2}{Department of Applied Mathematics and Theoretical Physics, University of Cambridge, Wilberforce Road, Cambridge CB3 0WA, UK}
\altaffiltext{3}{Graduate School of Advanced Integrated Studies in Human Survivability, Kyoto University, Kyoto 606-8501 Japan}
\altaffiltext{4}{Kwasan and Hida Observatories, Kyoto University, 17-1 Kitakazan-ohmine-cho, Yamashina-ku, Kyoto, 607-8471, Japan}
%%\email{shibata@kwasan.kyoto-u.ac.jp, isobe@kwasan.kyto-u.ac.jp}


\maketitle

\begin{abstract}

The recent observations of \citet{singh2012} have shown multiple plasma ejections and the intermittent
nature of magnetic reconnection in the solar chromosphere, highlighting the need for fast reconnection to occur in highly collisional plasma.
However, the physical process through which fast magnetic reconnection occurs in partially ionized plasma, like the solar chromosphere, is still poorly understood. It has been shown that for sufficiently high magnetic Reynolds numbers, Sweet--Parker current sheets can become unstable leading to tearing mode instability and plasmoid formation, but when dealing with a partially ionized plasma the strength of coupling between the ions and neutrals plays a fundamental role in determining the dynamics of the system. We propose that as the reconnecting current sheet thins and the tearing instability develops, plasmoid formation passes through strongly, intermediately, and weakly coupled (or decoupled) regimes, with the time scale for the tearing mode instability depending on the frictional coupling between ions and neutrals. We present calculations for the relevant time scales for fractal tearing in all three regimes. We show that as a result of the tearing mode instability and the subsequent non-linear instability due to the plasmoid-dominated reconnection, the Sweet--Parker current sheet tends to have a fractal-like structure, and when the chromospheric magnetic field is sufficiently strong the tearing instability can reach down to kinetic scales, which are hypothesized to be necessary for fast reconnection.
\end{abstract}

\section{Introduction}
Magnetic reconnection is a ubiquitous process where magnetic fields undergo topological rearrangement and which converts the magnetic energy into heat and various forms of plasma energy. Some of the key issues related with magnetic reconnection
are: 
\begin{enumerate}
\item How does fast magnetic reconnection occur and what determines the reconnection rate?
Two-dimensional magnetohydrodynamic (MHD) simulations of \textit{coronal} magnetic reconnection
show that the reconnection rate depends strongly on the resistivity model \citep{yokoyama1994}.
In the case of uniform resistivity, the reconnection becomes steady Sweet--Parker type, while in the case of anomalous resistivity, the reconnection tends to non-steady Petschek type. Apart from the
anomalous resistivity, the Hall effect also plays an important role in facilitating the fast reconnection.
\citep{oier01, cassak2005, cassak2007, drake08, ren08}. It is however not clear whether the Hall effect and anomalous resistivity operate and interact simultaneously in an astrophysical environment.

\item The anomalous resistivities required to explain the fast reconnection rate are $\textit{spatially}$ localized
\citep[e.g.][]{singh2007}. In this case, how can one reconcile the enormous gap between macroscopic scales
(e.g. solar flares $\sim 10^4$ km) and microscopic scales required to explain the reconnection rate?

\item How does magnetic reconnection occur in partially ionized, highly \textit{collisional} plasma?
The solar chromosphere is partially ionized and fully collsional. Anomalous resistivity appears in a collisionless plasma,
but, because of the incredibly small scales with which it is associated, it is quite unlikely that it will play a role in the solar chromosphere. The importance of partially ionized plasma on magnetic
reconnection in solar chromosphere is realized in some of the recent studies \citep{singh2011, zweibel2011, leake2012, leake2013, ni2015}. The multi-fluid MHD simulations of solar chromosphere have shown faster reconnection rates, compared to the single-fluid
Sweet--Parker prediction, and during the process of decoupling the recombination as well as the plasma outflow play a role in determining
the reconnection rate \citep{leake2012, leake2013}. The 2.5D MHD simulations of magnetic reconnection including the ambipolar diffusion and radiative cooling in the partially ionized solar chromosphere have been performed \citep{ni2015}. It has been found that fast magnetic
reconnection develops as a consequence of the plasmoid instability and there is no need to invoke anomalous resistivity.
\end{enumerate}

In this paper, we develop a non-linear model of magnetic reconnection in partially ionized plasma.
Recent observation of multiple plasma ejections in chromospheric anemone jets by \citet{singh2012} has shown a strongly time-dependent as well as the intermittent nature of magnetic reconnection in
the solar chromosphere. In this paper, we argue that as the magnetic reconnection in a Sweet--Parker current
sheet with a large aspect ratio goes through various regimes (strongly coupled, intermediate and weakly coupled or decoupled)
in a partially ionized plasma and the reconnection will be plasmoid-induced, fast and fractal-like.

\section{Magnetic Reconnection in Partially Ionized Plasma}
The solar chromosphere is a favorable site for magnetic reconnection to occur, since the Ohmic resistivity is greatest in that region--especially at the location of the temperature-minimum \citep{sturrock99}. In a partially ionized plasma, the collisions between electron-ion, electron-neutral and ion-neutral takes place. The electron-ion and electron-neutral collisions produce an Ohmic diffusion \citep{piddington54, cowling56, ni07, singh2010}. In addition to the Ohmic diffusion, the Hall drift and the ambipolar diffusion also appears in a partially ionized plasma. In the case of solar chromosphere, the Hall effect can often be ignored \citep{arber2009, singh2010, malyshkin2011}. The role of ambipolar diffusion has been studied analytically in the context of magnetic reconnection. It has been found that in a Sweet--Parker geometry, the ambipolar diffusion increases the reconnection rate \citep{vishniac1999, krishan2009}. Some of the observations of magnetic field in the solar chromosphere suggest that reconnection can occur at speeds $\sim$ 0.1 $V_{\rm A}$ \citep{dere96}. Recent high-resolution observations of jets in the solar chromosphere show a strongly time-dependent and intermittent nature of magnetic reconnection in the solar chromosphere. It was for the first time that \citet{singh2012}, based on the observations of multiple plasma ejections in the chromospheric anemone jets, suggested that if a time-dependent and bursty reconnection occurs 10 times longer than the Alfv\'en time scale, it can then naturally explain the observed intermittency of one to two minutes in the chromospheric jets. As a result of the ambipolar diffusion, the dissipation of currents is increased manifold compared to the standard Ohmic dissipation. The ambipolar diffusion can also contribute in the chromospheric heating and therefore the ambipolar diffusion should be included in the non-linear MHD simulations of chromospheric heating \citep{khomenko2012}.

It is known that in a partially ionized plasma, a current sheet undergoes thinning and enters a regime where the neutrals decouple from ions \citep{brand94, arber2009}. The strength of coupling between ion and neutrals in a partially ionized plasma can affect the reconnection rate and during the process of decoupling the reconnection rate could increase. The Alfv\'en speed in a decoupled regime is defined with respect to the ion mass density alone \citep{zweibel1989}. The multi-fluid MHD simulations show that, as a result of current sheet thinning and elongation, a critical Lundquist number ($S_{\rm critical}$) is reached in a partially ionized plasma and the plasmoid formation starts \citep{leake2012, leake2013}. During the current sheet thinning a stage is reached where the neutrals and ions decouple, and reconnection rate faster than the single-fluid Sweet--Parker prediction is observed in such multi-fluid simulations. The ion and neutral outflows are well coupled in the multi-fluid MHD simulations in the sense that the difference between ion and neutral
outflow is negligible compared to the magnitude of the ion outflow.

\section{Nonlinear and Intermittent Nature}
The breakup of Sweet--Parker current sheets with large aspect ratios and formation of plasmoids as a result of the tearing mode instability appears to be a generic feature of the reconnecting systems \citep{tanuma2001, shibata 2001, samtaney09, loureiro2007, loureiro2012}. There are number of theoretical studies on tearing mode instability that show the dependence of the growth rate of the instability on the ion-neutral collisions \citep{zweibel1989, zweibel2011}, the flows \citep{ofman91, loureiro2007}, the Hall drift \citep{baalrud11}, and the two-fluid effects \citep{mirnov04}. In this work, we focus on the role of ion-neutral collisions on tearing mode instability and fractal-like reconnection. The growth rate ($\gamma$) of tearing mode instability depends upon the extent of coupling and partial ionization in a partially ionized plasma as \citep{zweibel1989, zweibel2011}
\begin{equation}
\gamma^{\rm 5} (1 + \frac{\rho_{\rm n}}{\rho_{\rm i}} \frac{1}{1+ \frac{\gamma}{\nu_{\rm ni}}}) = \gamma_{\rm \star}^5,
\label{eq1}
\end{equation}
where $\gamma_{\rm \star}$ is the growth rate of tearing mode instability in a plasma where the Alfv\'en time is defined with respect to ions and the diffusion time includes Ohmic diffusivity, $\nu_{\rm ni}$ is the neutral-ion collision frequency, and
$\rho_{\rm n}$, $\rho_{\rm i}$ are the neutral and ion mass densities respectively.
For $\gamma/\nu_{\rm ni} \ll 1$, there exists a strongly coupled regime in a partially ionized gas. Apart from the strongly coupled regime, there exists an intermediate regime and a decoupled regime \citep{zweibel1989}. While deriving the dispersion relation in a partially ionized plasma [i.e. equation (1)], the effect of shear flows is not included. In general, we expect that various regimes would arise during the current sheet thinning in the presence of multi-fluid physics, ionization, recombination and heating processes. In what follows, we discuss the strongly coupled, intermediate and decoupled regimes in context of the nonlinear instability and fractal nature of the magnetic reconnection.

\subsection{Strongly Coupled Regime}
Equation (\ref{eq1}) in a strongly coupled regime reduces to
\begin{equation}
\gamma = \gamma_{\rm \star} f^{\rm 1/5}~,
\label{eq2}
\end{equation}
where $f = \frac{\rho_{\rm i}}{\rho_{\rm total}}$.
Lets consider the tearing mode instability in a Sweet--Parker current-sheet width that has a thickness and length of $\delta_{\rm n}$ and $\lambda_{\rm n}$ respectively, and $n$ in the subscript refers to
the $n$th-level tearing. So, the next level tearing will be $(n+1)$. For $n=0$, $\delta_{\rm 0}$
corresponds to the initial width of the current sheet.
Such a Sweet--Parker current sheet becomes unstable to the secondary tearing if
\begin{equation}
t_{\rm n} \leq \lambda_{\rm n} / V_{\rm A}
\label{eq3}
\end{equation}
where the timescale $t_{\rm n}$ refers to the the growth time of the tearing mode instability
occurring at maximum rate and it is given by
\begin{equation}
t_{\rm n} \simeq (t_{\rm diff} \tau_{\rm Ai})^{\rm 1/2} f^{\rm -1/5}~,
\label{eq4}
\end{equation}
where $V_{\rm Ai}$ is the Alfv\'en speed determined by the ion density and $V_{\rm A}$ is
the Alfv\'en speed determined by the total density as $V_{\rm A}=V_{\rm Ai}f^{\rm 1/5}$ and
$\tau_{\rm Ai} = \delta_{\rm n} / V_{\rm Ai}$. It was shown by \citet{zweibel1989} that by introducing $V_{\rm A}$,
the form of growth rate of tearing mode instability becomes similar to \citet{furth63}, hereafter FKR.
Equation (\ref{eq4}) can be simplified further to
\begin{equation}
t_{\rm n} \simeq (\frac{t_{\rm diff}}{\tau_{\rm Ai}})^{1/2} \tau_{\rm A}
\simeq (\frac{\delta_{\rm n}^{\rm 3} V_{\rm Ai}}{\eta})^{\rm 1/2} \frac{1}{V_{\rm A}}~,
\label{eq5}
\end{equation}
where $\tau_{\rm A} = \delta_{\rm n} / V_{\rm A}$.
The works by \citet{sonnerup81, biskamp86, biskamp92} can be referred for theory of the secondary tearing
in the Sweet--Parker Current Sheet. Remember here that $\lambda_{\rm n}/V_{\rm A}$ gives the time in which the perturbation is
carried out of the current sheet by the reconnection flow. So, we get
\begin{equation}
\delta_{\rm n} \leq \eta^{\rm 1/3} V_{\rm A}^{\rm -1/3} f^{\rm 1/6} \lambda_{\rm n}^{\rm 2/3}~.
\label{eq6}
\end{equation}
If the Eq.(\ref{eq6}) is satisfied during the strongly coupled regime, the secondary tearing of the Sweet--Parker current sheet occurs,
leading to the current sheet thinning in the nonlinear stage of the tearing mode instability. During this stage, the thickness of the
Sweet--Parker current sheet is determined by the most unstable wavelength ($\simeq 6 \delta_{\rm n} R_{\rm m*,n}^{\rm 1/4} $)
of the secondary tearing mode instability, i.e.,
\begin{equation}
\lambda_{\rm n+1} = 6 \eta^{\rm -1/4}V_{\rm Ai}^{\rm 1/4} f^{\rm 1/20} \delta_{\rm n}^{\rm 5/4}
\leq 6 \eta^{\rm 1/6}V_{\rm A}^{\rm -1/6} f^{\rm 5/24} \lambda_{\rm n}^{\rm 5/6}
\label{eq7}
\end{equation}
where $R_{\rm m*,n} = \delta_{\rm n} V_{\rm A} f^{\rm 1/5} / \eta$.
The critical wavelength for the tearing mode instability is an important deciding parameter here
\citep[See][]{shibata 2001}. The instability starts when the Sweet--Parker current sheet becomes longer than
the critical wavelength of tearing mode instability, given here by Eq. (\ref{eq7}). The current sheet will continue to
thin and when the current sheet thickness reaches
\begin{equation}
\delta_{\rm n+1} \leq \eta^{\rm 1/3} V_{\rm A}^{\rm -1/3} f^{\rm 1/6} \lambda_{\rm n+1}^{\rm 2/3}~,
\label{eq8}
\end{equation}
further tertiary tearing instability occurs.  Such subsequent tearing will occur again at a smaller scale.
It follows from Eqs. (\ref{eq6}) and (\ref{eq7}) that
\begin{equation}
\delta_{\rm n} \leq 6^{\rm 2/3} \eta^{\rm 1/6} (V_{\rm A})^{\rm -1/6} f^{\rm 1/6} \delta_{\rm n-1}^{\rm 5/6}~,
\label{eq9}
\end{equation}
or
\begin{equation}
\frac{\delta_{\rm n}}{L} \leq 6^{\rm 2/3} R_{\rm m}^{\rm -1/6} f^{\rm 1/6} (\frac{\delta_{\rm n-1}}{L})^{\rm 5/6}~,
\label{eq10}
\end{equation}
where
\begin{equation}
R_{\rm m} = \frac{L V_{\rm A} f^{\rm 1/5}}{\eta}~.
\label{eq11}
\end{equation}
It should be noted here that for $f=1$, the Eqs. (\ref{eq9}) to (\ref{eq11}) reduces to that of \citet{shibata 2001}.
The fractal tearing in strongly coupled regime continues till the current sheet thickness reaches
the microscopic scale of interest.
With a different $A$ i.e.
\begin{equation}\label{A_eq}
A = 6^{\rm 2/3} R_{\rm m}^{\rm -1/6} f^{\rm 1/6}~,
\label{eq12}
\end{equation}
the Eq.(9) can be rewritten as
\begin{equation}
\frac{\delta_{\rm n}}{L} \leq A^{\rm 6(1-(5/6)^{\rm n})} (\frac{\delta_{\rm 0}}{L})^{\rm (5/6)^{\rm n}}~.
\label{eq13}
\end{equation}

In the strongly coupled regime, we can estimate how many secondary tearings are required to reach
a length scale of $V_{\rm A}/\nu_{\rm ni}$. Here $V_{\rm A}$ is the bulk Alfv\'en speed given by
$V_{\rm A} = B/(4 \pi \rho_{\rm total})^{\rm 1/2}$ and $\nu_{\rm ni}$ is the neutral-ion collision frequency.
The ionization fraction is calculated from a photospheric-chromospheric model given in \citet{cox2000}. In the model
given in \citet{cox2000}, a hydrostatic equilibrium is assumed and the number densities in the solar atmosphere is
determined by solving the equations of radiative transfer and statistical equilibrium (without Local Thermodynamic Equilibrium).
The ionization fraction and the collision frequencies are plotted as a function of height in the solar atmosphere (FIGS. 1a,b).
The effect of non-equilibrium hydrogen ionization on the dynamical structure of the solar atmosphere and hydrogen line formation
is studied in the 2D simulations of \citet{leen07}. The simulations of \citet{leen07} show that the
chromospheric populations of the hydrogen (with principle quantum number = 2 level) are coupled to the ion populations and the
ionization fraction is also affected by the shock propagation in the solar chromosphere. The minimum ionization fraction ($\sim 10^{\rm -6}$)
in Fig. 1a turns out to be smaller than the non-equilibrium hydrogen ionization case (ionization fraction $\sim 10^{\rm -5}$).
Refer to \citet{singh2010} for the further details on the calculation of the collision frequencies
in the solar atmosphere. The magnetic field strength in the solar atmosphere is calculated using a power-law dependence
$B = B_{\rm 0} (\rho/\rho_{\rm s})^{\rm \alpha}$, where $B_{\rm 0}$ and $\rho_{\rm s}$ are the
magnetic field and the total mass density at the solar surface respectively. Figure 1c shows the variation of
magnetic field strength as a function of height and its dependence on the surface magnetic field and $\alpha$.
The neutral-ion collision frequency can be calculated from $\nu_{\rm ni} = (\rho_{\rm i}/\rho_{\rm n}) \nu_{\rm in}$. The ion-neutral collision frequency
($\nu_{\rm in}$) is given by $\nu_{\rm in} = n_{\rm n} \mu_{in} \sqrt{\frac{8 k_{\rm B} T}{\pi m_{\rm in}}} \Sigma_{\rm in}$ \citep[see ][]{khodachenko2004, singh2010}. Here $\Sigma_{\rm in} \sim 5 \times 10^{\rm -15}$~cm s$^{\rm -2}$ is the ion-neutral collision cross-section,
$\mu_{\rm in} = \frac{m_{\rm n}}{m_{\rm i}+m_{\rm n}}$ and $m_{\rm in} = \frac{m_{\rm i}m_{\rm n}}{m_{\rm i}+m_{\rm n}}$.
The length scale $V_{\rm A}/\nu_{\rm ni}$ is plotted as a function of height in the solar atmosphere (FIG. 2a).
For $V_{\rm A}$ = 10 kms$^{\rm -1}$ and $\nu_{\rm ni} = 10^{\rm 2}$ s$^{\rm -1}$, this length scale is about $10^{4}$~cm or $10^{-3} L$.
\begin{figure*}
\includegraphics[clip,width=55mm]{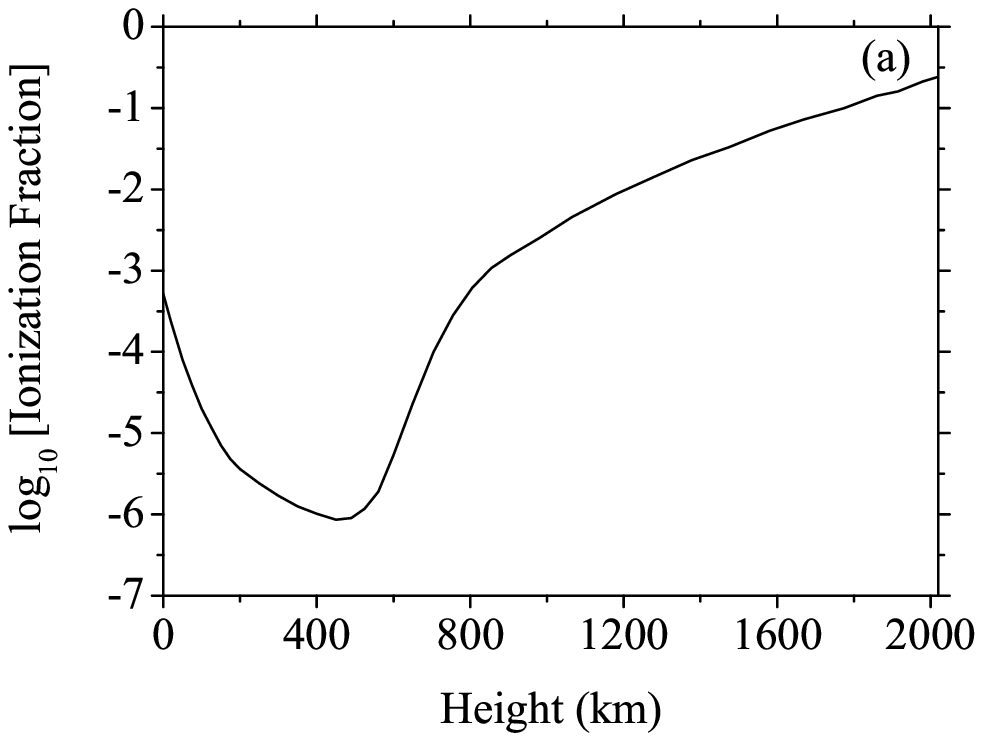}
\includegraphics[clip,width=55mm]{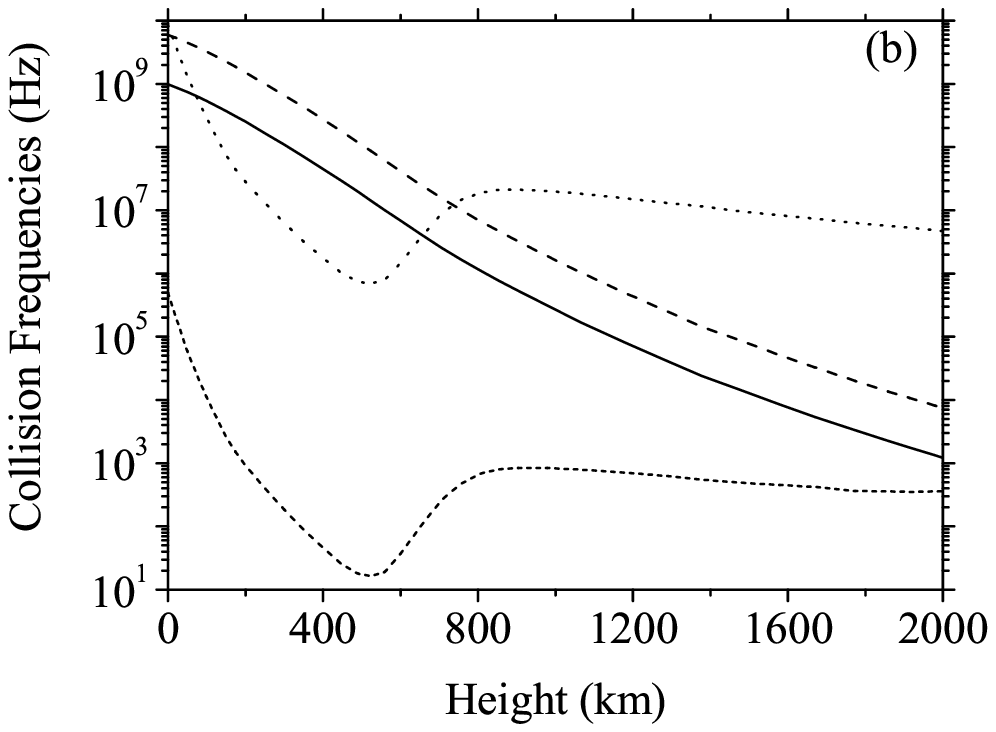}
\includegraphics[clip,width=55mm]{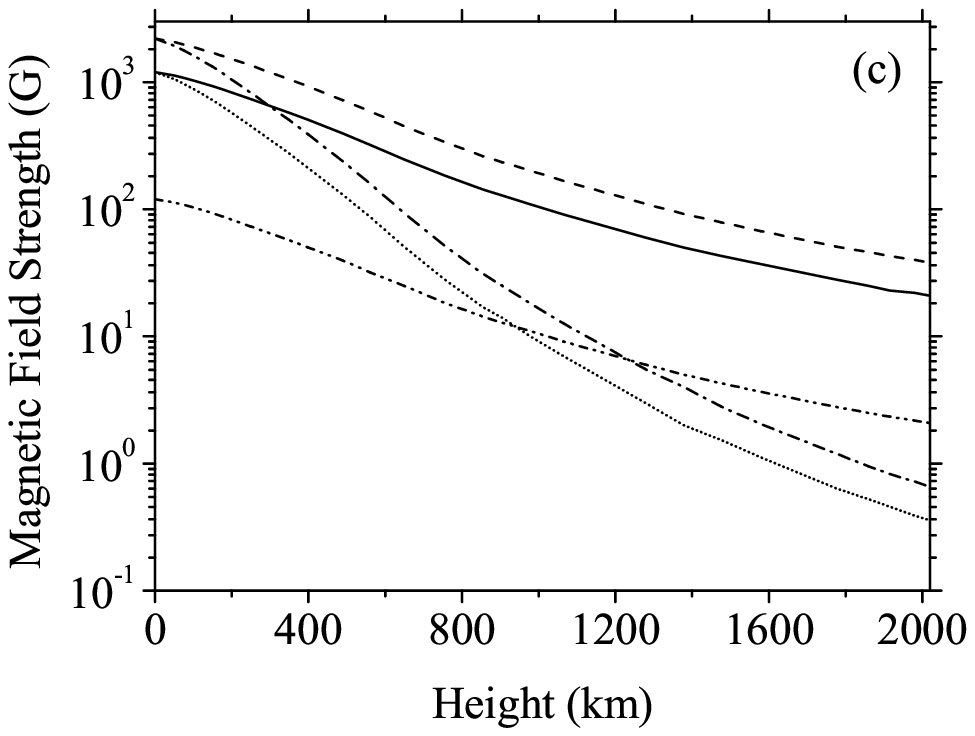}
\caption{(a) The ionization fraction i.e. $n_{\rm i}/(n_{\rm i}+n_{\rm n})$, (b)
the collision frequencies, and (c) the magnetic field strength as a function
of height in the solar atmosphere. In (b) the electron-ion collision frequency
is shown by a dotted line, the electron-neutral collision is shown by a dashed line, and the ion-neutral
collision frequency is shown by a solid line. In (c) the magnetic field strength for $B_{\rm 0} = 2.2$~kG
and $\alpha = 0.3$ is shown by a dashed line, for $B_{\rm 0} = 2.2$~kG
and $\alpha = 0.6$ is shown by a dash-dotted line, for $B_{\rm 0} = 1.2$~kG
and $\alpha = 0.3$ is shown by a solid line, $B_{\rm 0} = 1.2$~kG
and $\alpha = 0.6$ is shown by a short-dotted line, and for $B_{\rm 0} = 120$~G
and $\alpha = 0.3$ is shown by a dash-dot dotted line.}
\label{Fig1}
\end{figure*}
Taking the typical solar chromospheric values, $\delta_{\rm 0} = 10^{\rm 5}$~cm, $L=5 \times 10^{\rm 7}$~cm,
$\eta = 10^{\rm 7}$~ cm$^{\rm 2}$ s$^{-1}$, we find $R_{\rm m} = 10^{\rm 8}$ and A $\sim$ 0.05.
Just one tearing takes the current sheet to $3 \times 10^{\rm -4} L$.  We will see in the later sections that since
$V_{\rm Ai}/\nu_{\rm in}$ is of the order of $10^{\rm -4} L$, the plasma will shift from strong to decouple regime through the
intermediate state.
The time scale of n-th tearing ($t_{\rm n}$) in the strongly coupled regime is
\begin{equation}
t_{\rm n} \simeq (\frac{\delta_{\rm n}^{\rm 3} V_{\rm Ai}}{\eta})^{\rm 1/2}
\frac{1}{V_{\rm A}}~.
\label{eq14}
\end{equation}
Now, if we introduce $t_{\rm 0}$ as
\begin{equation}
t_{\rm 0} = \delta_{\rm 0}^{\rm 3/2} (V_{\rm Ai}^{\rm 1/2} \eta^{\rm -1/2} / V_{\rm A})~,
\label{eq15}
\end{equation}
and $A_{\rm 0} = 6^{\rm 2/3} R_{\rm m*,0}^{\rm -1/6} f^{\rm 1/6}$, and $R_{\rm m*,0}
= \delta_{\rm 0} V_{\rm A} f^{\rm 1/5} / \eta$,
then the scaling relations similar to \citet{shibata 2001} for
$\delta_{\rm n} / \delta_{\rm 0}$, $t_{\rm n}$ and $t_{\rm n} / t_{\rm n-1}$.
The time ($t_{\rm total}$) taken from $t_{\rm 1}$ to $t_{\rm n}$ for the solar chromospheric
conditions becomes
\begin{equation}
t_{\rm total} \leq (1.6 \times 10^{\rm -1}) t_{\rm 0}~.
\label{eq16}
\end{equation}

\subsection{Intermediate Regime}
For $\nu_{\rm ni} \ll \gamma \ll \nu_{\rm in}$, there exists an intermediate regime. The growth rate (Eq. \ref{eq1}) of tearing mode instability reduces to
\begin{equation}
\gamma = \gamma_{\rm \star}^{5/4} \nu_{\rm in}^{-1/4}~,
\label{eq17}
\end{equation} where $\nu_{\rm in}$ is the ion-neutral collision frequency.
The Sweet--Parker current sheet becomes unstable to secondary tearing if
\begin{equation}
t_{\rm n} \leq \lambda_{\rm n} / V_{\rm Ai}.
\label{eq18}
\end{equation}
Here we use rather restrictive condition because on the large scales the plasma can be coupled.
The growth time of the tearing instability at maximum rate ($t_{\rm n}$) and it is given by
\begin{equation}
t_{\rm n} \simeq (t_{\rm diff} t_{\rm Ai})^{\rm 5/8} \nu_{\rm in}^{1/4}
\simeq (\frac{\delta_{\rm n}^{\rm 3}}{\eta V_{\rm Ai}})^{\rm 5/8} \nu_{\rm in}^{1/4}~.
\label{eq19}
\end{equation}
In the intermediate regime, the time for the reconnection flow to carry the perturbation out of the current sheet
is $\lambda_{\rm n}/V_{\rm Ai}$,
we get
\begin{equation}
\delta_{\rm n} \leq \eta^{\rm 1/3} V_{\rm Ai}^{\rm -1/5} \nu_{\rm in}^{\rm -2/15} \lambda_{\rm n}^{\rm 8/15}~.
\label{eq20}
\end{equation}
The most unstable wavelength of the secondary tearing instability is given by
\begin{equation}
\lambda_{\rm n+1}
= 6 \eta^{\rm -1/4}V_{\rm Ai}^{\rm 1/4} \delta_{\rm n}^{\rm 5/4}
\leq 6 \eta^{\rm 1/6} \nu_{\rm in}^{\rm -1/6}  \lambda_{\rm n}^{\rm 2/3}
\label{eq21}
\end{equation}
While deriving Eq.(27), $R_{\rm m*,n} = \delta_{\rm n} V_{\rm Ai} / \eta$ is taken.
As the current sheet thickness arrives at
\begin{equation}
\delta_{\rm n+1} \leq \eta^{\rm 1/3} V_{\rm Ai}^{\rm -1/5} \nu_{\rm in}^{\rm -2/15} \lambda_{\rm n+1}^{\rm 8/15},
\label{eq22}
\end{equation}
further tertiary tearing occurs. It follows from Eqs. (20) and (21) that
\begin{equation}
\delta_{\rm n} \leq 6^{\rm 8/15} \eta^{\rm 1/5} (V_{\rm Ai})^{\rm -1/15} \nu_{\rm in}^{\rm -2/15} \delta_{\rm n-1}^{\rm 2/3}~,
\label{eq23}
\end{equation}
or
\begin{equation}
\frac{\delta_{\rm n}}{L} \leq A (\frac{\delta_{\rm n-1}}{L})^{\rm 2/3}~,
\label{eq24}
\end{equation}
where
\begin{equation}
A = 6^{\rm 8/15} R_m^{\rm -1/5} \tau_{\rm Ai}^{-2/15} \nu_{\rm in}^{\rm -2/15}~,
\label{eq25}
\end{equation}
and
\begin{equation}
R_{\rm m} = \frac{L V_{\rm Ai}}{\eta}~.
\label{eq26}
\end{equation}
The Eqs. (23) and (24) are new equations in terms of scalings and they are fundamentally different from
\citet{shibata 2001}. The Eq.(23) further leads to
\begin{equation}
\frac{\delta_{\rm n}}{L} \leq A^{\rm 3(1-x)} (\frac{\delta_{\rm 0}}{L})^{\rm x}
\label{eq27}
\end{equation}
where
\begin{equation}
x = (2/3)^{\rm n}~.
\label{eq28}
\end{equation}
For the current sheet thinning in the intermediate regime, we can estimate how many secondary tearings would be required to reach to a scale
where the reconnection approaches a decoupling scale. The length scale of the decoupling is given by
\begin{equation}
L_{\rm dec} = \frac{V_{\rm Ai}}{\nu_{\rm in}} = \frac{B}{(4 \pi m_{\rm i} n_{\rm i})^{\rm 1/2} n_{\rm n} \mu_{in} \Sigma_{\rm in}}
(\frac{8 k_{\rm B} T}{\pi m_{\rm in}})^{\rm -1/2}.
\label{eq29}
\end{equation}
Substituting the typical numbers in the solar chromosphere,
\begin{equation}
L_{\rm dec} \simeq 400 (\frac{B}{30 \textup{G}}) (\frac{n_{\rm i}}{10^{\rm 11} \textup{cm}^{\rm -3}}\frac{T}{10^{\rm 4} \textup{K}})^{\rm -1/2}
(\frac{n_{\rm n}}{10^{\rm 13} \textup{cm}^{\rm -3}})^{\rm -1} \textup{cm}.
\label{eq30}
\end{equation}
For $V_{\rm Ai}$ = 300~ kms$^{\rm -1}$ and $\nu_{\rm in} = 10^{\rm 5}$~s$^{\rm -1}$, $L_{\rm dec}$ = 300 cm.
The Alfv\'en velocity w.r.t. ions could be 10 times higher and then the $L_{\rm dec} = 3 \times 10^{\rm 3}$~cm
(about 10$^{\rm -4} L$). If we use $t_{\rm n}\nu_{\rm in}=1$ from Eq.(24) and calculate $\delta_n/L$ then we get
$\delta_{\rm decouple} = (\eta V_{\rm Ai} \nu_{\rm in}^{\rm -2})^{\rm 1/3}$.

\begin{figure*}
\includegraphics[clip,width=90mm]{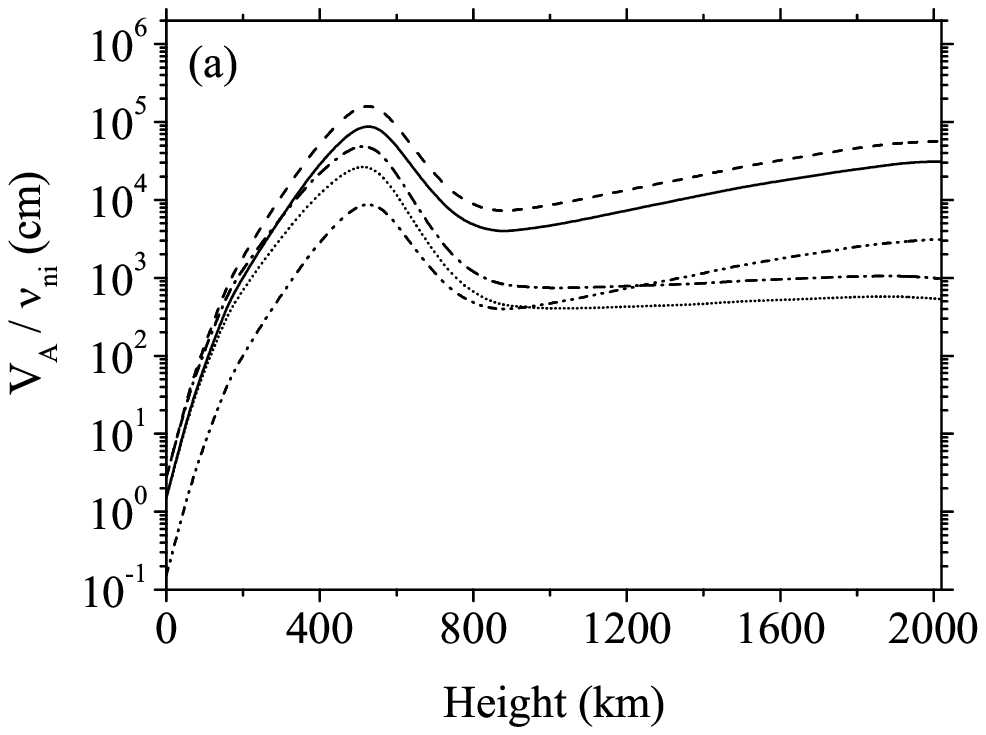}
\includegraphics[clip,width=90mm]{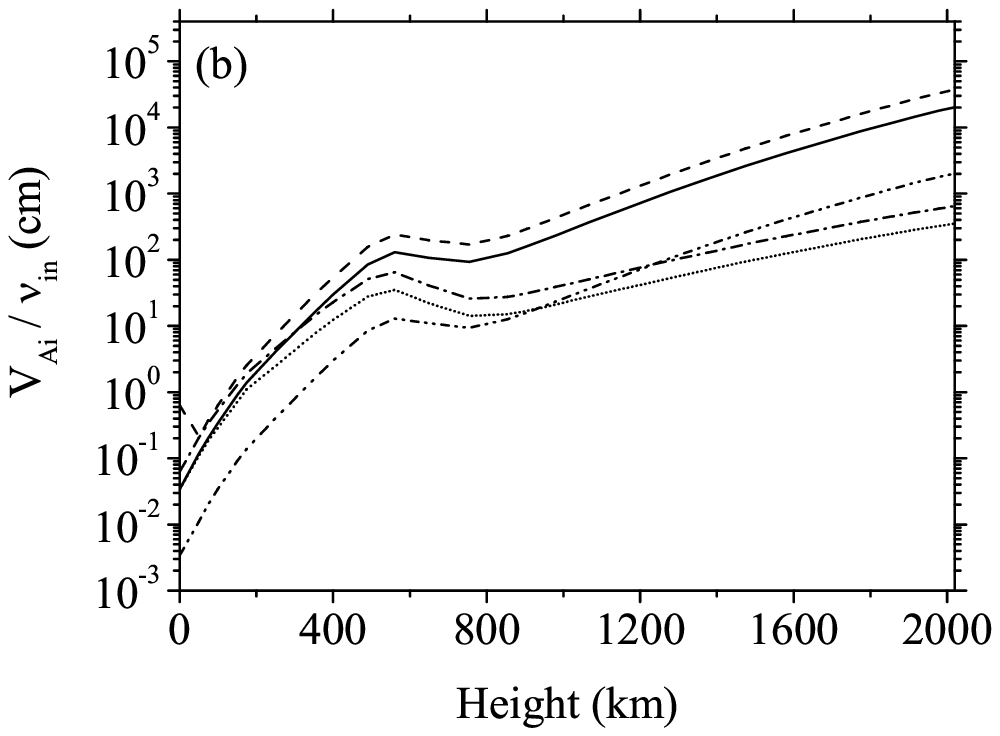}
\caption{(a) The variation of length-scales $V_{\rm A}/\nu_{\rm ni}$, and (b) $V_{\rm Ai}/\nu_{\rm in}$ in cm.
as a function of height in the solar atmosphere. Here $V_{\rm A}$ is the bulk Alfv\'en speed. The length-scales
are calculated for different values of $\alpha$ and the magnetic field strength $B_{\rm 0}$ at the solar surface.
The length-scales for $B_{\rm 0} = 2.2$~kG and $\alpha = 0.3$ is shown by a dashed line, for $B_{\rm 0} = 2.2$~kG
and $\alpha = 0.6$ is shown by a dash-dotted line, for $B_{\rm 0} = 1.2$~kG
and $\alpha = 0.3$ is shown by a solid line, $B_{\rm 0} = 1.2$~kG
and $\alpha = 0.6$ is shown by a short-dotted line, and for $B_{\rm 0} = 120$~G
and $\alpha = 0.3$ is shown by a dash-dot dotted line.}
\label{Fig2}
\end{figure*}

The variation of ($V_{\rm Ai}/{\nu_{\rm in}}$) is plotted as a function of height in the solar atmosphere (FIG. 2b).
Taking the typical solar chromospheric values, $\delta_{\rm 0} = 10^{\rm 5}$~cm, $L=5 \times 10^{\rm 7}$~cm,
$\eta = 10^{\rm 7}$~ cm$^{\rm 2}$ s$^{-1}$, we find $R_{\rm m} = 10^{\rm 8}$ and A $\sim$ 0.09.
Since $\delta_{\rm n}$ should be smaller than the typical decoupling scale ($L_{\rm dec}$) , we have
\begin{equation}
\frac{\delta_{\rm n}}{L} < \frac{L_{\rm dec}}{L}~,
\label{eq31}
\end{equation}
or
\begin{equation}
(0.1)^{\rm 3(1-(2/3)^{\rm n})} (2 \times 10^{\rm -3})^{\rm (2/3)^{\rm n}} < 10^{\rm -4}
\label{eq32}
\end{equation}
This inequality gives that in about four secondary tearings ($n=4$), the $\delta_{\rm n}$ approaches 10$^{\rm -4} L$.
The time scale of n-th tearing is
\begin{equation}
t_{\rm n} \simeq (\delta_{\rm n}^{\rm 3}/(\eta V_{\rm Ai}))^{\rm 5/8} \nu_{\rm in}^{\rm 1/4}~,
= (\frac{\delta_{\rm n}}{\delta_{\rm 0}})^{\rm 15/8} t_{\rm 0}~,
\label{eq33}
\end{equation}
where
\begin{equation}
t_{\rm 0} = \delta_{\rm 0}^{\rm 15/8} (\eta V_{\rm Ai})^{\rm -5/8} \nu_{\rm in}^{\rm 1/4}~.
\label{eq34}
\end{equation}
Eqs. (33) and (34) lead to
\begin{equation}
\frac{\delta_{\rm n}}{\delta_{\rm 0}} \simeq A_{\rm 0}^{(1-(2/3)^{\rm n})}
\label{eq35}
\end{equation}
where $A_{\rm 0} = 6^{\rm 8/5} R_{\rm m*,0}^{\rm -3/5} \tau_{\rm Ai}^{\rm -2/5} \nu_{\rm in}^{\rm -2/5}$,
and $R_{\rm m*,0}
= \delta_{\rm 0} V_{\rm Ai} / \eta$ and $\tau_{\rm Ai} = \delta_{\rm 0} / V_{\rm Ai}$, we find
\begin{equation}
t_{\rm n} \simeq A_{\rm 0}^{15/8(1-(2/3)^{\rm n})} t_{\rm 0}~.
\label{eq36}
\end{equation}
Thus we obtain
\begin{equation}
t_{\rm n} / t_{\rm n-1} = A_{\rm 0}^{\rm (5/8)(2/3)^{(n-1)}} \leq A_{\rm 0}^{\rm 1/2}
\label{eq37}
\end{equation}
for $n \ge 1$. It follows from this equation that
\begin{equation}
t_{\rm n} \leq A_{\rm 0}^{\rm 5/8} t_{\rm n-1} \leq A_{\rm 0}^{\rm (5n/8)} t_{\rm 0}.
\label{eq38}
\end{equation}
Using Eq. (38), we can get the total time $t_{\rm total}$ as
\begin{equation}
t_{\rm total} \leq t_{\rm 0} A_{\rm 0}^{\rm 5/8}
 \frac{1-A_{\rm 0}^{\rm 5n/8}}{1-A_{\rm 0}^{\rm 5/8}} \leq t_{\rm 0} A_{\rm 0}^{\rm 5/8}.
\label{eq39}
\end{equation}
Now the $t_{\rm total}$ can be calculated taking the typical solar chromospheric conditions
and $\delta_{\rm 0} = 10^{\rm 4}$~cm which is of the order of $V_{\rm A}/\nu_{\rm ni}$. We get
\begin{equation}
t_{\rm total} \leq (9.4 \times 10^{-1}) t_{\rm 0}.
\label{eq40}
\end{equation}
For $\delta_{\rm 0} = 10^{\rm 5}$~cm, we get
$t_{\rm total} \leq (3 \times 10^{-2}) t_{\rm 0}$.
It is important to note that the time-scale $t_{\rm 0}$ appearing in Eq. (40) is given by Eq. (34).
\subsection{Decoupled Regime}
At this stage the current sheet width as reached to
a stage where $\delta_{\rm 0} = L_{\rm dec}$.
In the decoupled regime ($\gamma \gg \nu_{\rm in}$), the Eq. (1) reduces to
$\gamma = \gamma_{\rm \star}$. The scaling relations derived by \citet{shibata 2001} can be used by taking the Alfv\'en speed w.r.t ions i.e.
 \begin{equation}
\delta_{\rm n} \leq 6^{\rm 2/3} (\frac{\eta}{V_{\rm Ai}})^{\rm 1/6} \delta_{\rm n-1}^{\rm 5/6}~,
\label{eq41}
\end{equation}
or
\begin{equation}
\frac{\delta_{\rm n}}{L} \leq 6^{\rm 2/3} R_{\rm m}^{\rm -1/6} (\frac{\delta_{\rm n-1}}{L})^{\rm 5/6}~,
\label{eq42}
\end{equation}
where
\begin{equation}
R_{\rm m} = \frac{L V_{\rm Ai}}{\eta}.
\label{eq43}
\end{equation}
Then Eq. (41) leads to
\begin{equation}
\frac{\delta_{\rm n}}{L} \leq A^{\rm 6(1-(5/6)^{\rm n})} (\frac{\delta_{\rm 0}}{L})^{\rm }.
\label{eq44}
\end{equation}
For the typical solar chromospheric values mentioned earlier, this gives $A (= 6^{\rm 2/3} R_{\rm m}^{\rm -1/6})$ $\sim$ 0.15.
After ten secondary tearings, the current sheet width reaches to a scale where
$\delta_{\rm n} \sim 10^{\rm -5} L$.
Taking $t_{\rm 0}$ as
\begin{equation}
t_{\rm 0} = \delta_{\rm 0}^{\rm 3/2} / (\eta V_{\rm Ai})^{\rm 1/2}~,
\label{eq46}
\end{equation}
and using the scaling relation for $t_{\rm n}$ from \citet{shibata 2001}, we get
$t_{\rm total}$ as
\begin{equation}
t_{\rm total} \leq 6 \times 10^{\rm -1} t_{\rm 0}.
\label{eq48}
\end{equation}

\section{Results}
\begin{figure*}
\includegraphics[clip,width=80mm]{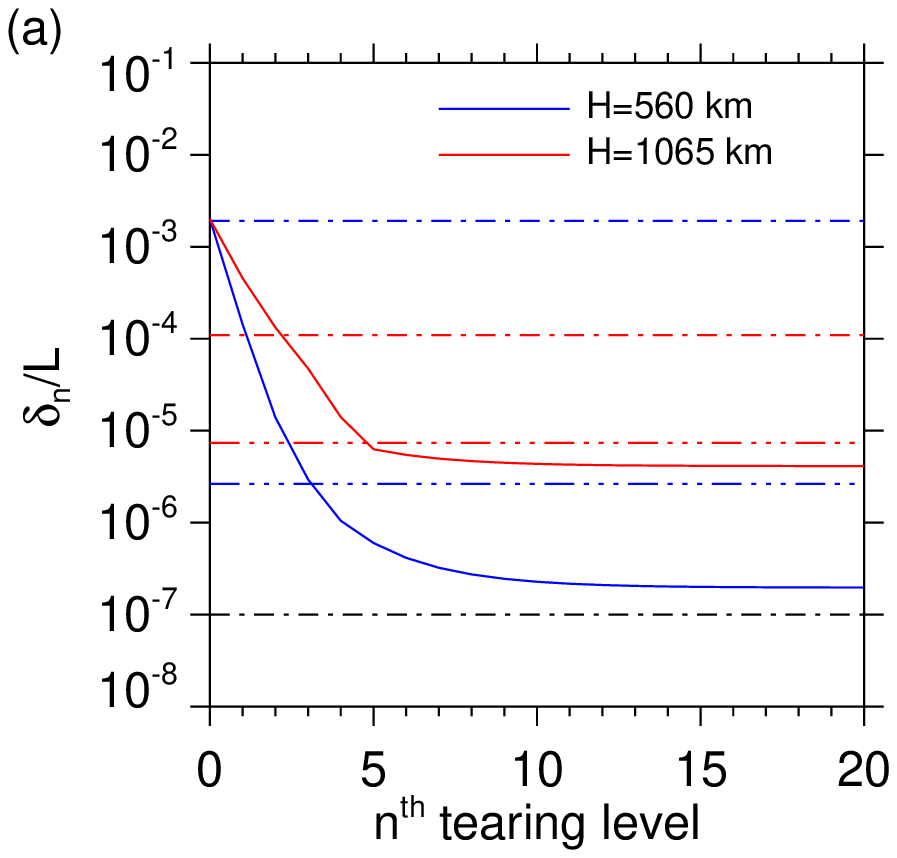}
\includegraphics[clip,width=80mm]{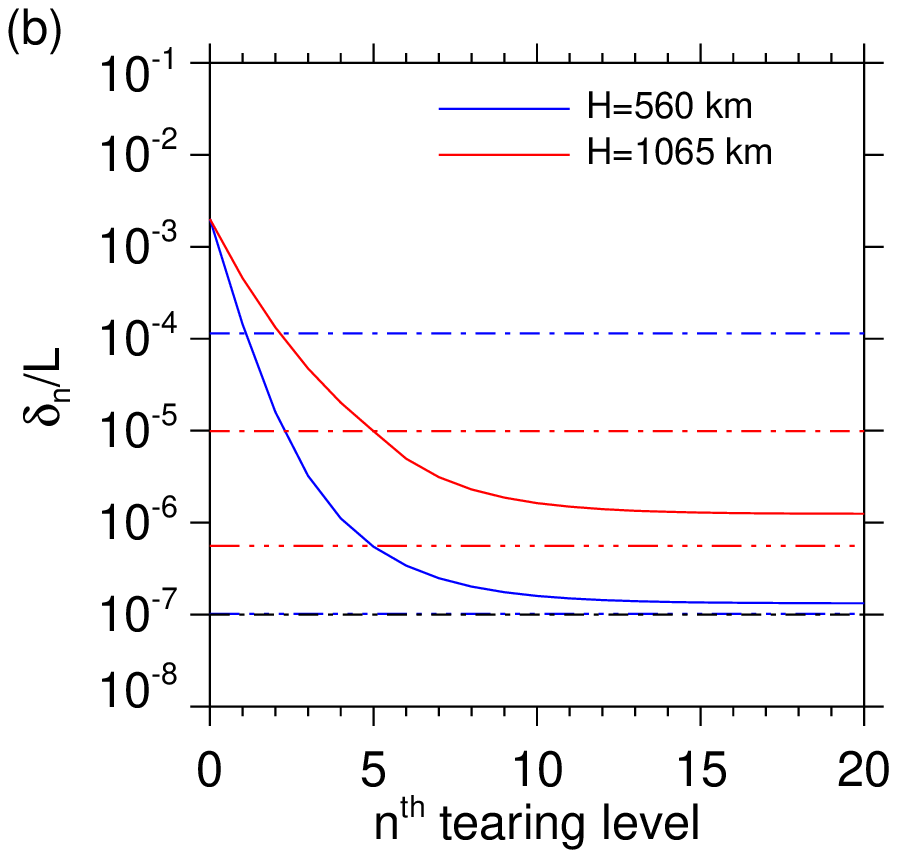}\\
\includegraphics[clip,width=80mm]{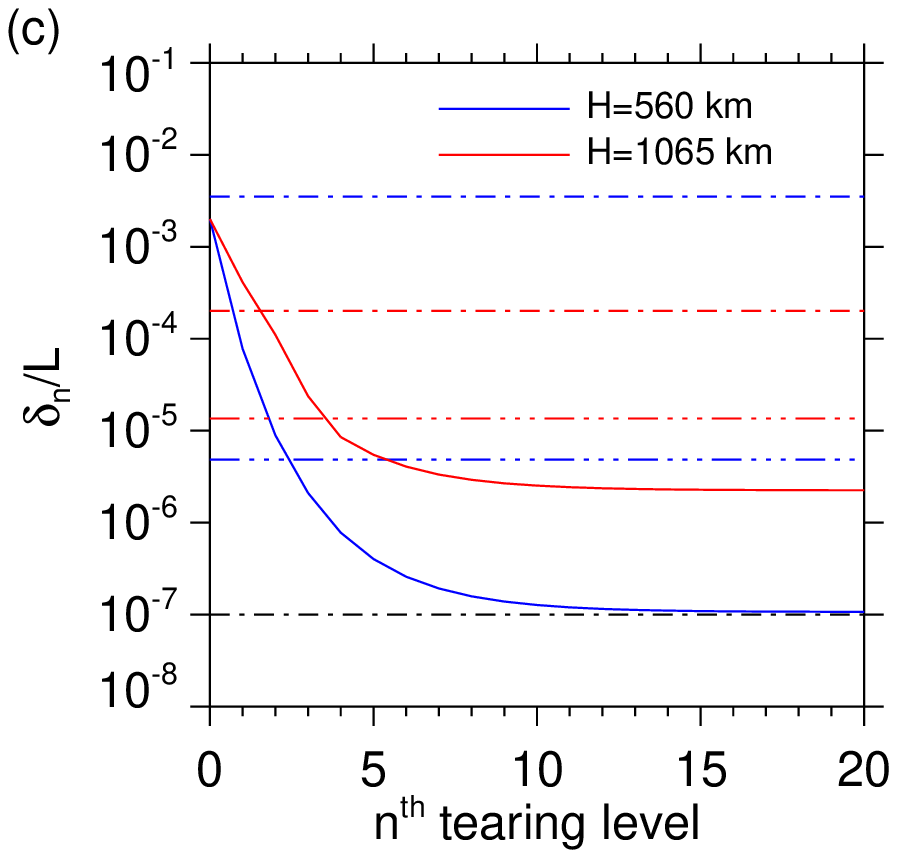}
\includegraphics[clip,width=80mm]{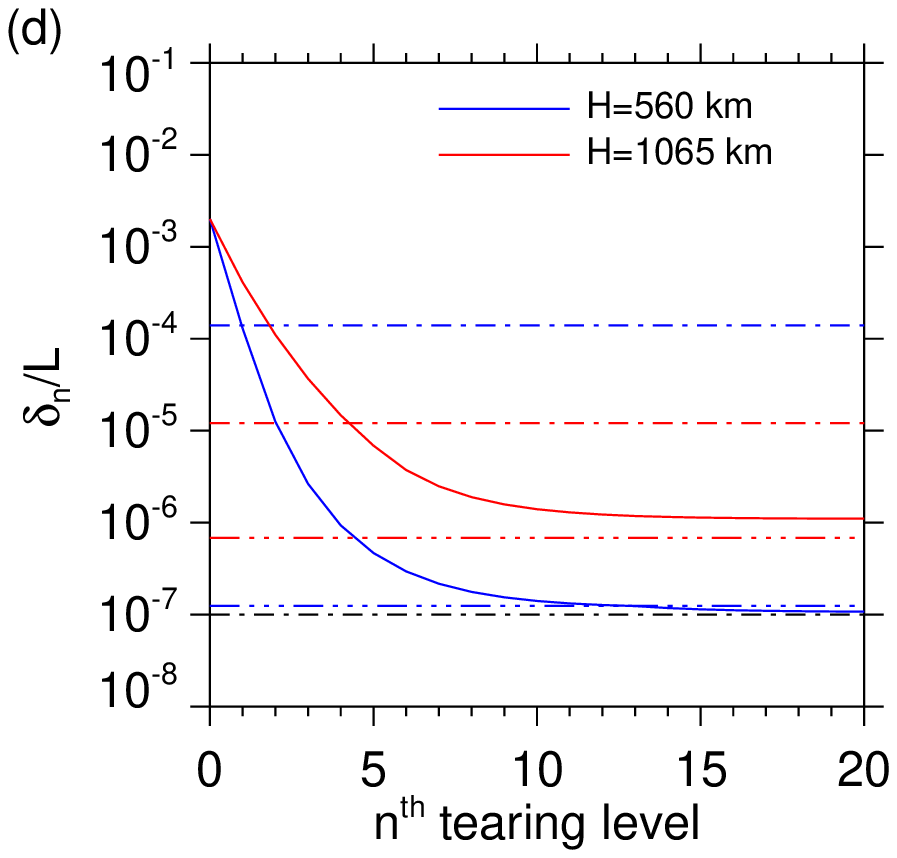}
\caption{The normalised current sheet width at the nth level of tearing where
we consider the decoupling lengthscale to be determined by the Alfv\'en velocity (a and c) and the tearing timescale (b and d) for the magnetic field models of 1.2 kG with $\alpha=0.3$ (a and b) and 2.2 kG with $\alpha=0.3$ (c and d).
The blue and red solid lines give the evolution for the atmospheric parameters calculated at the heights $560$\,km and $1065$\,km respectively. The dot-dash horizontal lines show the decoupling lengthscales (coloured)
and the ion Larmor radius (black).}\label{Fig3}
\end{figure*}

Figure \ref{Fig3} shows the width of the current sheet normalised by the length of
the current sheet at the n-th level of tearing for two different assumptions for the decoupling length
scale using different models for the magnetic field and different heights in the solar atmosphere.
For panels (a) and (c) we assume that the decoupling length scale is determined by $L_{\rm dec}=V_{\rm Ai}/\nu_{\rm in}$ and for Panels (b) and (d) we take $L_{\rm dec}=(\eta V_{\rm Ai} \nu_{\rm in}^{\rm -2})^{\rm 1/3}$.
The key difference between these two is the second has smaller decoupling lengths, which makes it harder to tear down
to the decoupling length scale.

For the case shown in panels (a) and (c), the decoupling length scale is sufficiently large that for both cases
under study the decoupling length scale is reached. This happens in under 5 tearings for all cases shown here.
However, it is only the case with the strongest magnetic field where the system reaches down to the kinetic scales.

For the case shown in panels (b) and (d), the decoupling length scale is much smaller so
that in only the case with the strongest magnetic field where the decoupling length scale is
reached. This happens in on the 12th-tearing, after which the system approaches the kinetic scales.
In all cases it is very interesting to note that for the lower ion Alfv\'en velocities (implying either higher ionisation fraction or a lower magnetic field strength) the fast reconnection regime is not achieved.

It is possible to determine if a system can reach the the kinetic scale purely by assessing the asymptotic limit of Eq. (\ref{eq44}), i.e.:
\begin{equation}\label{limit}
\frac{\delta_{\rm \infty}}{L} = A^{\rm 6}
\label{eq49}
\end{equation}
Therefore, to reach the kinetic scales found at $10^{\rm -7}L$, we require $A \le 0.0681$ or a Lundquist number of $R_{\rm m}\ge 1.3 \times 10^{\rm 10}$.
For the resistivity and current sheet length used in the paper we require an ion Alfv\'en velocity of $V_{\rm Ai} = 2.6\times 10^{\rm 9}$\,cm\,s$^{\rm -1}$ and $\eta = 10^{\rm 7}$\,cm$^{\rm 2}$\,s$^{\rm -1}$. This naturally places a strong constraint on the particular chromospheric conditions where fast reconnection can occur, with the general implication that a strong magnetic field is a necessary requirement to achieve this.
It is also important to note the importance of the factor $6^{\rm 2/3}$ found in A (see Eq. \ref{A_eq}) plays an important role in this limit, i.e. the factor is equal to $6^{\rm 4}$.
In other works, e.g. \citet{sturrock94}, this factor is taken as $2\pi/1.4 \sim 4.5$ which would result in a factor of $3$ reduction in the limit given by Equation \ref{limit}.

\section{Current sheet width}

In this analysis, we have assumed that the initial current sheet width is given by the
Sweet--Parker scaling. However, when we consider ion-neutral drift, it is possible that
the current sheet will undergo thinning as a result of the neutrals leaving the current
sheet before reconnection takes place. In fact, this drift velocity in the current sheet
was estimated to be approximately $1$~kms$^{\rm -1}$ by \citet{singh2011}.
There are a number of simple ways with which the new current sheet width can be estimated,
and we provide a few of these here.

At first we try to find out the width of the current sheet at the point where the
ambipolar diffusion balances the Ohmic diffusion.
This is given by the following equation:
\begin{equation}
\eta\mathbf{J}=\frac{B(L_{\rm 0})^2}{\nu_{\rm in}\rho_{\rm i}}\mathbf{J}
\label{eq50}
\end{equation}
for some distance from the centre of the current sheet ($L_{\rm 0}$).
\citet{brand94} showed that the magnetic field takes the distribution of $B\propto x^{\rm 1/3}$, therefore we have:
\begin{equation}
\eta=\frac{L_{\rm 0}^{\rm 2/3}}{\nu_{\rm in}\rho_{\rm i}}
\label{eq51}
\end{equation}
Implying that
\begin{equation}
L_{\rm 0}=(\eta\nu_{\rm in}\rho_{\rm i})^{\rm 3/2}
\label{eq52}
\end{equation}
This new current sheet width $L_{\rm 0}$ can be seen as analogous to the new current sheet width found for the Kippenhahn-Schl\"{u}ter prominence model by \citet{hillier2010}.

Another possibility is that there exists a weak guide field component in the centre of the current sheet.
In this case the ambipolar diffusion squeezes the current sheet until this guide field becomes strong enough
to support it, this was numerically investigated by \citet{arber2009}.
In this case we are looking for a balance between the external magnetic field ($B_{\rm ex}$) and the guide field after it has undergone
compression to $B_{\rm newg}$. Through flux conservation, the original guide field ($B_{\rm g}$) roughly relates to ($B_{\rm newg}$) in the following way:
\begin{equation}
L_{\rm cs}B_{\rm g}=l_{\rm cs} B_{\rm newg}
\label{eq53}
\end{equation}
where $L_{\rm cs}$ is the original width of the current sheet.
As we need the point were $B_{\rm ex}=B_{\rm newg}$ we get:
\begin{equation}
l_{\rm cs} = L_{\rm cs} (\frac{\rm B_g}{B_{\rm ex}})
\label{eq54}
\end{equation}
This estimate can be seen as consistent with the thinning found in the numerical experiments as shown in Figure 12 of \citet{arber2009}.

It is interesting to note that the nature of the equation for the current sheet width at the n-th tearing
(Eqs. 13, 27 and 44 ) implies that we do not have to be overly concerned about the initial width of the current
sheet from the point of view of whether or not we can have fast reconnection.
As eq. 33, in the limit of $x \rightarrow \infty$, becomes $\delta_{\rm \infty}/L = A^{\rm 3}$, which is
independent of the initial current sheet width. To state this simply, reaching the decoupling length scale
is independent of the initial current sheet width. It does, however, play a role in determining how many levels
of tearing are necessary to reach this width, and so may be important for speeding up the process of reaching
the decoupling length scale.

\section{Critical Lundquist Number}
The multi-fluid, two-dimensional MHD simulation in a Harris type current sheet shows fast reconnection and plasmoid formation in
the solar chromosphere \citep{leake2012, leake2013}. It is noticed in the simulations that as a result of current sheet thinning and elongation
, the plasmoid instability in the MHD simulation starts when the aspect ratio (= $\lambda_{\rm n}/\delta_{\rm n}$) approaches 200 and the
$S_{\rm critical} \sim 10^{\rm 4}$ \citep{leake2012}. \citet{biskamp86} also noticed that for sufficiently long current sheets the tearing mode
becomes unstable despite of the stabilizing effect of the inhomogeneous flow. In an intermediate regime in partially ionized plasma, the maximum growth rate of tearing mode instability is given by Eq. (17). This relation is based upon the FKR-theory but gets modified due to the ion-neutral collisions. The Sweet--Parker current sheet becomes unstable to tearing mode when
\begin{equation}
t_{\rm n} < \lambda_{\rm n}/V_{\rm Ai}~.
\label{eq55}
\end{equation}
From Eq.(18), we get
\begin{equation}
(\frac{\delta_{\rm n}^{\rm 3}}{\eta V_{\rm Ai}})^{\rm 5/8} \nu_{\rm in}^{1/4}
< \lambda_{\rm n}/V_{\rm Ai}~.
\label{eq56}
\end{equation}
The Eq. (53) can be re-written as
\begin{equation}
(\frac{\lambda_{\rm n} V_{\rm Ai}}{\eta})^{\rm 3/8}
< \lambda_{\rm n}^{11/8} \delta_{\rm n}^{\rm -15/8}
\eta^{\rm 1/4} \nu_{\rm in}^{-1/4}~,
\label{eq57}
\end{equation}
and that further reduces to
\begin{equation}
R_{\rm ms}^{\rm 3/8} <
 A_{\rm aspect}^{\rm 11/8} \delta_{\rm n}^{\rm -1/2} \eta^{\rm 1/4} \nu_{\rm in}^{\rm -1/4}~,
\label{eq58}
\end{equation}
where $R_{\rm ms} = (\lambda_{\rm n} V_{\rm Ai}/\eta)$ and $A_{\rm aspect}$ is the aspect
ratio of the Sweet--Parker current sheet given by $A_{\rm aspect}=(\lambda_{\rm n}/\delta_{\rm n})$.
The Sweet--Parker reconnection rate ($M_{\rm SP}$) is given by $M_{\rm SP} = R_{\rm ms}^{\rm -1/2} = A_{\rm aspect}^{\rm -1}$.
This along with Eq. (54) gives
\begin{equation}
M_{\rm SP}^{\rm -3/4} = A_{\rm aspect}^{\rm 3/4}
< A_{\rm aspect}^{11/8} \delta_{\rm n}^{\rm -1/2} \eta^{\rm 1/4} \nu_{\rm in}^{\rm -1/4}~.
\label{eq59}
\end{equation}
For the secondary tearing to occur the critical aspect ratio should be
\begin{equation}
A_{\rm aspect} \sim \delta_{\rm n}^{\rm 4/5} \eta^{\rm -2/5} \nu_{\rm in}^{\rm 2/5}~.
\label{eq60}
\end{equation}
This gives $A_{\rm aspect} \sim 250$ if we consider $\delta_{\rm n} = \delta_{\rm 0} = 10^{\rm 4}$~cm,
$\nu_{\rm in} = 10^{\rm 5}$~s$^{\rm -1}$, $\eta = 10^{\rm 7}$~ cm$^{\rm 2}$ s$^{-1}$. The current-sheet
width used in Eq. (57) is considered in the intermediate regime. Since $\lambda_{\rm n} = \eta^{\rm -1} V_{\rm Ai} \delta_{\rm n}^{\rm 2}$,
the value of $S_{\rm critical}$
\begin{equation}
S_{\rm critical} = \frac{\lambda_{\rm n} V_{\rm Ai}}{\eta} = A_{\rm aspect}^{\rm 2} = 6.3 \times 10^{\rm 4}~.
\label{eq61}
\end{equation}
It should be noted that the $S_{\rm critical}$ obtained here is similar to the value reported in multi-fluid MHD simulations. There are
even higher values of $A_{\rm aspect}$ that are reported in the simulations of \citet{leake2013}. While noticing that the $S_{\rm critical}$
depends upon the Prandtl number as well \citet{loureiro2013} it is important to mention here that if we take slightly different values of
$\eta = 10^{\rm 6.5}$~ cm$^{\rm 2}$ s$^{-1}$ and $\nu_{\rm in} = 10^{\rm 6}$~s$^{\rm -1}$ and keep the current sheet width unchanged in Eq. (57),
we get $A_{\rm aspect} \sim 1000$ and $S_{\rm critical} \sim 10^{\rm 6}$.

\section{Discussions}
The magnetic reconnection, as it proceeds from a coupled regime to a decoupled regime, undergoes rapid changes in the tearing mode time scale. Once the tearing mode instability develops, the plasmoids are formed. The role of two-dimensional flows on the current sheet and plasmoid formation is studied by \citet{loureiro2007}. The maximum growth rate of tearing mode in \citet{loureiro2007} scales with Lundquist number
as $\gamma_{\rm max} \sim R_{\rm m}^{\rm 1/4}(V_{\rm A}/L)$ where $R_{\rm m} = L V_{\rm A}/\eta$, and the corresponding wave number scales as
$k_{\rm max} \sim R_{\rm m}^{\rm 3/8}$. It should be noted here that the scalings based on FKR theory (without including effect of shear flow) on Sweet--Parker current sheet leads to the scaling laws that are consistent with \citet{loureiro2007}, so a simple application of \citet{zweibel1989} would provide correct answer to the problem. The formation of plasmoids and its ejection from the long and thin current sheet lead to a nonlinear instability and fractal like reconnection. The current sheet thickness ($\delta_{\rm n}/L$) is shown in the n-th secondary tearing for the intermediate and the decoupled regimes. The relevant time scales of fractal tearing are also calculated for an intermediate coupling and decoupled case in partially ionized plasma. The tearing mode time scale ($t_{\rm 0}$) could lie between 10 s - 150 s \citep[c.f.][]{nishizuka2011}.
This gives $t_{\rm total} = 0.6 t_{\rm 0}$ between 6 s - 90 s in the decoupled regime. It is clear that although the main energy release is
explained by the fast reconnection, the plasmoid induced instability and fractal tearing can produce a very thin current sheet in a partially
ionized plasma. The shift in the regimes from strongly coupled to weakly coupled is one of the important component in the fast magnetic reconnection in partially ionized plasma. As a result of the plasmoid mediated processes in a partially ionized plasma, there could be few more regimes that will bring the scale down to ion-skin depth or ion-Larmor radius due to the Hall or kinetic effects \citep[e.g.][]{daughton09, shepherd10}.

As is clear from Figure \ref{Fig3}, it is not necessarily the case that reconnection occurring in a partially ionized plasma will reach the kinetic scale.
In fact, we have been able to estimate the required Lundquist number (based on the ion Alfv\'en velocity) for kinetic scales to be reached as $R_{\rm m} \ge 1.3 \times 10^{10}$, i.e. this is the $R_{\rm m}$ required to allow the tearing instability to still be important down to kinetic scales.
When it is possible to reach these small scales due to multiple tearings, the reconnection has to pass through three distinct regimes: the coupled regime, the intermediate regime and the decoupled regime. As the decoupling of the Alfv\'en waves happens before the decoupling of the instability, it can be expected that the reconnection dynamics undergoes multiple transitions. It is worth noting that there is another scale that we have not considered here, the Hall lengthscale, which is related to the lengthscale at which ions become demagnetised. We plan to investigate how the inclusion of Hall physics influences the cascade we describe in a future work.

The passage through reconnection regimes will be characterised by the development of the decoupling of the neutrals from the reconnection dynamics, first by decoupling from the plasma then the plasma decoupling from the neutrals. This does not mean that that these neutrals are never to feel the ions again, on the contrary the ions will recouple to the neutrals and even then later the neutrals will recouple to the plasma as the reconnected plasma is ejected from the current sheet and starts an inverse cascade, partially through plasmoid merger, to connect to the global reconnection dynamics. This coupling process will allow for very short periods of time, protons and electrons to reach very high temperatures ($\propto V_{\rm Ai}^{\rm 2}$) before they collisionally lose their heat to the neutrals. It is possible that high frequency observations of chromospheric reconnection would reveal the presence of high energy thermal emission (potentially in $\gamma$-rays) at the reconnection site.

In this paper, we have presented a mechanism for reaching kinetic scales through reconnection in a partially ionized plasma. The interesting problem that
we have addressed here is important to the basic physics of magnetic reconnection in the solar chromosphere, and could potentially be applied in other
partially ionized, astrophysical plasmas. However, there is still one question that we must address: Is significant flux reconnected at small scales?
This is a point of great interest as if only a small amount of flux is being reconnected quickly, it is unlikely that the global reconnection rate would greatly differ. To discuss this, we use the results from two studies performed on the plasmoid instability in a fully ionized plasma.
\citet{huang2012} gives the distribution of the frequency of plasmoids with a given flux as $f(\psi)\propto\psi^{\rm -1}$ and \citet{uzdensky2010} gives $f(\psi)\propto\psi^{\rm -2}$. For the former, this implies that the total flux reconnected at any scale should be the same as any other scale, the latter implies that the reconnected flux is dominated by small scales with a $\psi^{\rm -1}$ distribution. Therefore, this leads us to believe that significant flux is reconnected at small scales, allowing the cascade to collisionless reconnection that our model describes to present a significant increase in the reconnection rate.

\bigskip
AH is supported by his STFC Ernest Rutherford Fellowship grant number ST/L00397X/1, and for part of this work by the Grant-in-Aid for Young Scientists (B, 25800108).
Part of this work was performed during a visit by KAPS to Kyoto University, during this visit KAPS was supported by the Grant-in-Aid for Young Scientists (B, 25800108).
The authors would also like to thank the anonymous referee for their helpful and insightful comments.

\end{document}